\documentclass[a4paper,twocolumn,11pt,accepted=2017-05-09]{quantumarticle}
\pdfoutput=1
\usepackage[utf8]{inputenc}
\usepackage[english]{babel}
\usepackage[T1]{fontenc}
\usepackage{amsmath}
\usepackage{hyperref}

\usepackage{tikz}
\usepackage{lipsum}

\begin{document}

\title{A brain-inspired paradigm for scalable quantum vision}

\author{Chenghua Duan}
\affiliation{School of Computer science and Technology, Beijing Institute of Technology, 100081 Beijing, China}
\orcid{0009-0000-6427-7218}

\author{Xiuxing Li}
\affiliation{School of Computer science and Technology, Beijing Institute of Technology, 100081 Beijing, China}

\author{Wending Zhao}
\affiliation{Huayi Boao (Beijing) Quantum Technology Co., Ltd, 100081 Beijing, China}

\author{Lin Yao}
\affiliation{Huayi Boao (Beijing) Quantum Technology Co., Ltd, 100081 Beijing, China}

\author{Qing Li}
\affiliation{School of Computer science and Technology, Beijing Institute of Technology, 100081 Beijing, China}

\author{Ziyu Li}
\affiliation{School of Computer science and Technology, Beijing Institute of Technology, 100081 Beijing, China}

\author{Fukang Li}
\affiliation{School of Computer science and Technology, Beijing Institute of Technology, 100081 Beijing, China}

\author{Junhao Ma}
\affiliation{School of Computer science and Technology, Beijing Institute of Technology, 100081 Beijing, China}

\author{Xia Wu}
\affiliation{School of Computer science and Technology, Beijing Institute of Technology, 100081 Beijing, China}
\orcid{0000-0002-2377-6093}
\email{wuxia@bit.edu.cn}

\maketitle

\begin{abstract}
One of the fundamental tasks in machine learning is image classification, which serves as a key benchmark for validating algorithm performance and practical potential.
However, effectively processing high-dimensional, detail-rich images, a capability that is inherent in biological vision, remains a persistent challenge.
Inspired by the human brain's efficient ``Forest Before Trees'' cognition, we propose a novel Guiding Paradigm for image recognition, leveraging classical neural networks to analyze global low-frequency information and guide targeted quantum circuit towards critical high-frequency image regions.
We present the Brain-Inspired Quantum Classifier (BIQC), implementing this paradigm via a complementarity architecture where a quantum pathway analyzes the localized intricate details identified by the classical pathway.
Numerical simulations on diverse datasets, including high-resolution images, show the BIQC's superior accuracy and scalability compared to existing methods.
This highlights the promise of brain-inspired, hybrid quantum-classical approach for developing next-generation visual systems.
\end{abstract}

\section{Introduction}

Decoding visual information is a key challenge in machine learning, with profound implications across diverse domains ranging from security to healthcare~\cite{grossman2019convergent, shen2017deep}.
Early algorithms mainly relied on feature extraction and Bayesian-based matching methods but struggled with deformations and generalization~\cite{lin2011large, boiman2008defense, kutsuna2012active}.
The introduction of deep learning, particularly convolutional neural networks (CNNs), revolutionized the field by enabling the automatic learning of hierarchical feature representations~\cite{lecun1995comparison}.
The breakthrough work by Krizhevsky et al., known as AlexNet, showcased the power of deep CNNs, leading to substantial performance improvements in image classification tasks~\cite{krizhevsky2012imagenet}.
Recently, innovations like generative models and data augmentation have endowed image recognition algorithms with stronger generalization capabilities~\cite{chen2024towards, mikolajczyk2018data}.
However, the current deep learning algorithms often struggle with accuracy when dealing with large, high-dimensional, and detail-rich images~\cite{jiao2019survey, xu2023comprehensive, gupta2024pushing}.
Moreover, high computational cost is another major challenge in the era of big data~\cite{samsi2023words, zheng2024llamafactory}.

Insights from biological computation, particularly neuroscience, continue to drive significant innovation in artificial intelligence (AI)~\cite{hassabis2017neuroscience,ullman2019using,lee2012neural}.
Brain-inspired AI, exemplified by the Neocognitron model~\cite{fukushima1980neocognitron}, has already achieved significant advancements.
This foundational neural network for pattern recognition was explicitly designed by Kunihiko Fukushima based on Hubel and Wiesel's research on hierarchical feature extraction in the visual cortex~\cite{hubel1959receptive}.
Notably, humans tend to perceive the overall structure of a scene (the ``forest'') before the local details (the ``trees''), known as ``Forest before Trees'', which was originally observed and experimentally documented by cognitive psychologist David Navon~\cite{ref32, ref33}.
This mechanism reduces cognitive load and maintains accuracy without compromising processing efficiency~\cite{ref40}.
Inspired by this, we propose an innovative Guiding Paradigm: the classical network assesses the complete ``forest'', and the quantum circuit focusing on ``trees''.

Quantum computing offers a novel paradigm potentially suited for analyzing such complex, detailed information (the ``trees''), leveraging the properties of quantum information for potentially exponential computational speedup~\cite{nielsen2010quantum, rieffel2011quantum, horowitz2019quantum}.
Parameterized quantum circuits represent a leading approach for near-term Quantum Machine Learning (QML) applications~\cite{schuld2015introduction,huang2018demonstration,cong2019quantum}, and combining quantum modules with classical networks has shown promise in image processing~\cite{gawron2020multi,henderson2020quanvolutional,sebastianelli2021circuit}.
However, applying quantum computing directly, especially in the context of big data, faces significant challenges including limited qubit counts, fidelity issues, barren plateaus, and difficulties mapping large classical datasets to quantum states~\cite{mcclean2018barren,montanaro2016quantum,cerezo2022challenges}.
Quantum-classical hybrid computing, integrating quantum components within classical architectures, thus emerges as a pragmatic solution to harness quantum capabilities while mitigating these limitations~\cite{adhikari2022hybrid,endo2021hybrid}.

Current hybrid approaches predominantly follow parallel~\cite{ref31} or serial~\cite{henderson2020quanvolutional, sebastianelli2021circuit, ref35, ref36} paradigms.
While these models have demonstrated performance advantages and promise across various tasks, they often exhibit drawbacks.
Specifically, parallel architectures can suffer from inefficiencies and information loss during data conversion, while serial architectures typically demand extensive quantum resources unsuitable for high-dimensional inputs~\cite{ref31, ref22, ref38}.
Thus, developing a new hybrid architecture that enhances the synergy between quantum and classical components and overcomes these specific issues is crucial.

In this work, we present the Brain-Inspired Quantum Classifier (BIQC), a novel architecture embodied with the Guiding Paradigm.
BIQC employs a dual-channel design that mimics the complementary low spatial frequency (LSF) and high spatial frequency (HSF) processing in human vision~\cite{ref23, ref24, ref25, ref26}.
The classical pathway efficiently extracts global features and identifies HSF regions, while the quantum pathway performs detailed analysis of local patches.
Through extensive experiments across diverse, high-resolution image datasets, including natural scenes, medical images, and texture datasets, BIQC outperforms current classical and QML methods.
This brain-inspired Guiding Paradigm offers a resource-efficient and scalable path for the wide application of QML in real-world tasks.

\begin{figure*}[!t]
    \centering
    \includegraphics[width=0.96\linewidth]{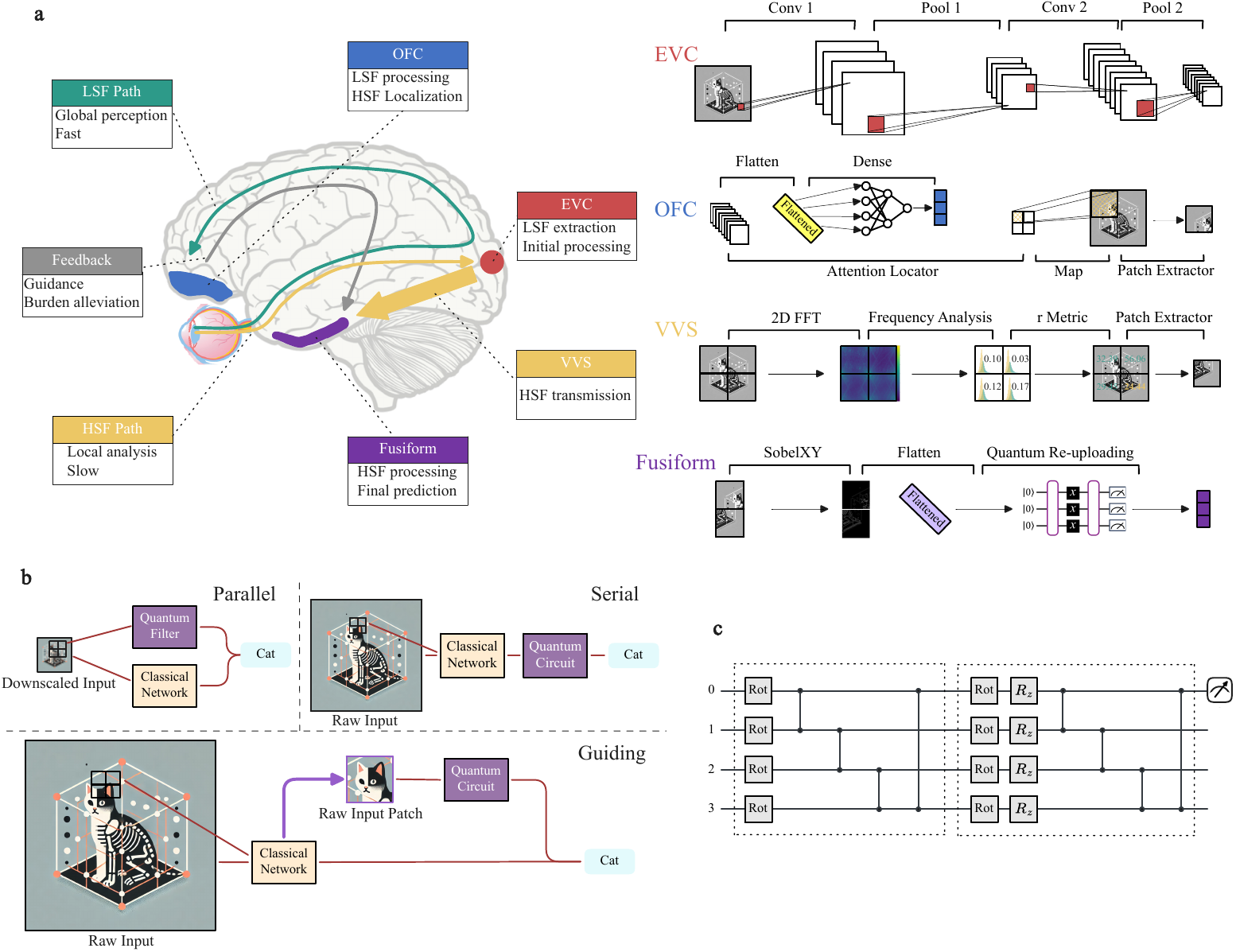}
    \caption{\textbf{BIQC architecture and Guiding Paradigm for scalable QML.}
    \textbf{a}, Overall architecture of the BIQC model, a dual-pathway framework mimicking human visual processing. The classical pathway, simulating the EVC and OFC, rapidly extracts LSF features and localizes potential HSF regions using an attention mechanism. Complementarily, the quantum pathway, inspired by the VVS and Fusiform gyrus, utilizes a 2D-DFT metric to precisely identify HSF regions and employs a data re-uploading quantum circuit for detailed recognition of these critical patches. Outputs from both pathways are integrated for final classification.
   \textbf{b}, Hybrid quantum-classical paradigms for image classification: Parallel, Serial, and Guiding. In the Guiding Paradigm, inspired by human cognition, a classical network guides a quantum circuit to focus on complex regions of the raw input, contrasting with Parallel (quantum and classical filters process raw input simultaneously) and Serial (quantum component processes hidden states from classical network) paradigms.
   \textbf{c}, Circuit diagram of a representative 4-qubit data re-uploading ansatz with CZ gate entanglement, illustrating the quantum circuit design within the BIQC's Fusiform module.}
    \label{fig:main}
\end{figure*}

\section{Methods}

\subsection{Brain-inspired QML paradigms}

Based on the idea that humans perceive coarse structures of a scene before scrutinizing local details~\cite{ref32}, we develop a quantum-classical Guiding Paradigm aimed at high-dimensional image classification.
The traditional quantum-classical hybrid computing paradigms typically employ standard parallel or serial frameworks, as shown in Fig.~\ref{fig:main}(b).
Our paradigm assigns specific roles to classical and quantum modules, complementing each other.
The classical pathway quickly processes global LSF signals and identifies high-complexity regions.
These regions are then sent to a quantum circuit focusing on finer details, which reduces the dimensionality load on the quantum circuit.

For instance, in the $96 \times 96$ STL-10~\cite{coates2011analysis} bird vs. plane classification task, the Guiding Paradigm effectively localizes a $32 \times 32$ HSF patch encompassing the bird's rectrices (tail feathers).
Detailed analysis of this patch reveals subtle textures, such as the organic texture of feathers compared to the metallic surface of an airplane rudder, which might be overlooked by traditional networks focusing solely on global features.
This targeted approach allows for efficient allocation of quantum resources to the most discriminative local regions, thereby enhancing classification accuracy.

By confining quantum operations to HSF-rich patches, our paradigm capitalizes on quantum representational strengths without incurring the excessive resource demands commonly seen in high-resolution tasks.
This design mitigates qubit limitations and minimizes information loss from repeated data transformations.

\subsection{The implementation of BIQC}
\label{subsec:methods_BIQC}

As illustrated in Fig.~\ref{fig:main}(a), our algorithm inspired by human visual system research~\cite{ref23, ref24, ref25}, includes four steps: early visual cortex (EVC), orbitofrontal cortex (OFC), ventral visual stream (VVS), and Fusiform gyrus.
(i) EVC: Quickly processes visual inputs to extract low-frequency information.
(ii) OFC: Generates initial predictions based on LSF features and identifies potential HSF regions in the original input, which mimics the human brain's attention.
(iii) VVS: Introduces a metric for selecting HSF regions that contain high-frequency feature information of the image.
(iv) Fusiform: Enhances identified regions given by VVS and OFC and uses quantum circuits for pattern recognition.

Neuroscience research indicates that the human brain utilizes a complex dual-channel system for image recognition~\cite{ref24, ref23, ref25, ref26}.
Similarly, the BIQC algorithm employs a dual-channel approach for efficient image recognition: 
(i) Classical pathway (LSF channel): Mimicking EVC and OFC, this pathway extracts global structural cues from input images like contours and shapes, and pinpointing regions potentially rich in critical HSF details.
(ii) Quantum pathway (HSF channel): Analogous to the VVS culminating in the fusiform gyrus, this channel employs a data re-uploading quantum circuit to analyze local patches, processing high-frequency cues like edges and textures in the quantum domain~\cite{ref06}.
This design strategically directs computationally demanding quantum resources to intricate image regions, addressing scalability limitations inherent in conventional hybrid QML approaches.

The translation of these neuro-inspired principles into the BIQC algorithm's architecture involves several key computational stages across its classical and quantum pathways, which are detailed below.

Mimicking the EVC's role in efficient initial processing to capture coarse structure, the classical pathway utilizes a CNN to extract global, LSF features.
This CNN comprises two convolutional layers and max-pooling operations, rapidly processing input images $\mathbf{X}$ to generate LSF features $\mathbf{Z}_{\mathrm{LSF}}$.
Analogous to the OFC's guidance of attention, an attention map $\mathbf{A}_{\mathrm{HSF}}$ is then derived from $\mathbf{Z}_{\mathrm{LSF}}$ via a $1 \times 1$ convolutional layer to highlight potential HSF regions $\mathbf{P}^{\mathrm{Human}}_{\mathrm{HSF}}$ for targeted quantum processing.

HSF regions are identified through two complementary strategies.
One approach selects regions via the attention mechanism, $\mathbf{P}^{\mathrm{Human}}_{\mathrm{HSF}}$.
The second, critical for focusing on intricate details, draws inspiration from the VVS, which excels at identifying and transmitting HSF signals from complex visual areas.
We implement this by employing a 2D-DFT based metric ($r$-metric) to quantify image patch complexity.
Specifically, the input image is first divided into non-overlapping regions $R_{i,j}$ of size $h \times h$.
For each region $R_{i,j}$, we compute the 2D-DFT coefficients $C_{u,v}$ as:
\begin{equation}
  C_{u,v} = \sum_{x=0}^{h-1} \sum_{y=0}^{h-1} R_{i,j}(x,y) e^{-2\pi i \left( \frac{ux}{h} + \frac{vy}{h} \right)},
\end{equation}
where $x$ and $y$ index pixel coordinates within the region, and $u$ and $v$ are frequency indices.
We then define a frequency magnitude $f_{u,v} = \sqrt{\left( \frac{u}{h} \right)^2 + \left( \frac{v}{h} \right)^2 }$ and separate the frequency coefficients into low ($\mathcal{S}_{\mathrm{Low}}$) and high ($\mathcal{S}_{\mathrm{High}}$) frequency sets based on a cutoff frequency $f_c$.
The $r$-metric for region $R_{i,j}$ is finally calculated as the ratio of energies in high and low frequencies:
\begin{equation}
    r_{R_{i,j}} = \sqrt{ \frac{E_{\mathrm{High}}}{E_{\mathrm{Low}}} } = \sqrt{ \frac{\sum_{(u,v) \in \mathcal{S}_{\mathrm{High}}} |C_{u,v}|^2}{\sum_{(u,v) \in \mathcal{S}_{\mathrm{Low}}} |C_{u,v}|^2} },
\end{equation}
The patch $\mathbf{P}^{\mathrm{Metric}}_{\mathrm{HSF}}$ with the maximal $r$-metric is then selected for quantum processing.

The quantum pathway emulates the Fusiform gyrus's role in detailed HSF analysis for recognition.
It employs a data re-uploading quantum circuit (Fig.~\ref{fig:main}(c)) to analyze localized HSF patches, aiming to capture the complex, high-frequency patterns crucial for classification.
This data re-uploading circuit, denoted as $f_{\mathrm{Quantum}}$, consists of $L$ stacked NonlinearQuantumBlocks.
Each block, $f^{(l)}$, applies a parameterized quantum circuit $U^{(l)}(\mathbf{z}^{(l)}_n, \boldsymbol{\theta}^{(l)})$ with data re-uploading, followed by Pauli Z measurements to generate a quantum feature vector $\mathbf{h}_{\mathrm{Quantum}}$. The circuit $U^{(l)}$ is structured as:
\begin{equation}
  \begin{aligned}
    U^{(l)}(\mathbf{z}^{(l)}_n, \boldsymbol{\theta}^{(l)}) & = \\
    \prod_{k=1}^L \left[ U_{SE}(\boldsymbol{\theta}^{(l)}_k) \cdot \bigotimes_{j=1}^d R_z(z^{(l)}_{n,j}) \right] & \cdot U_{SE}(\boldsymbol{\theta}^{(l)}_{out}),
  \end{aligned}
\end{equation}
where $U_{SE}(\boldsymbol{\theta})$ represents a strongly entangling layer, and $R_z(z^{(l)}_{n,j})$ encodes classical data $z^{(l)}_{n,j}$ through Z-axis rotation gates.

Features extracted from both classical and quantum pathways are concatenated and fed into a fully classical connected layer with Sigmoid activation for binary classification, integrating LSF global context with detailed HSF local features for efficient and scalable image recognition.

\begin{figure*}[!t]
    \centering
    \includegraphics[width=1.0\linewidth]{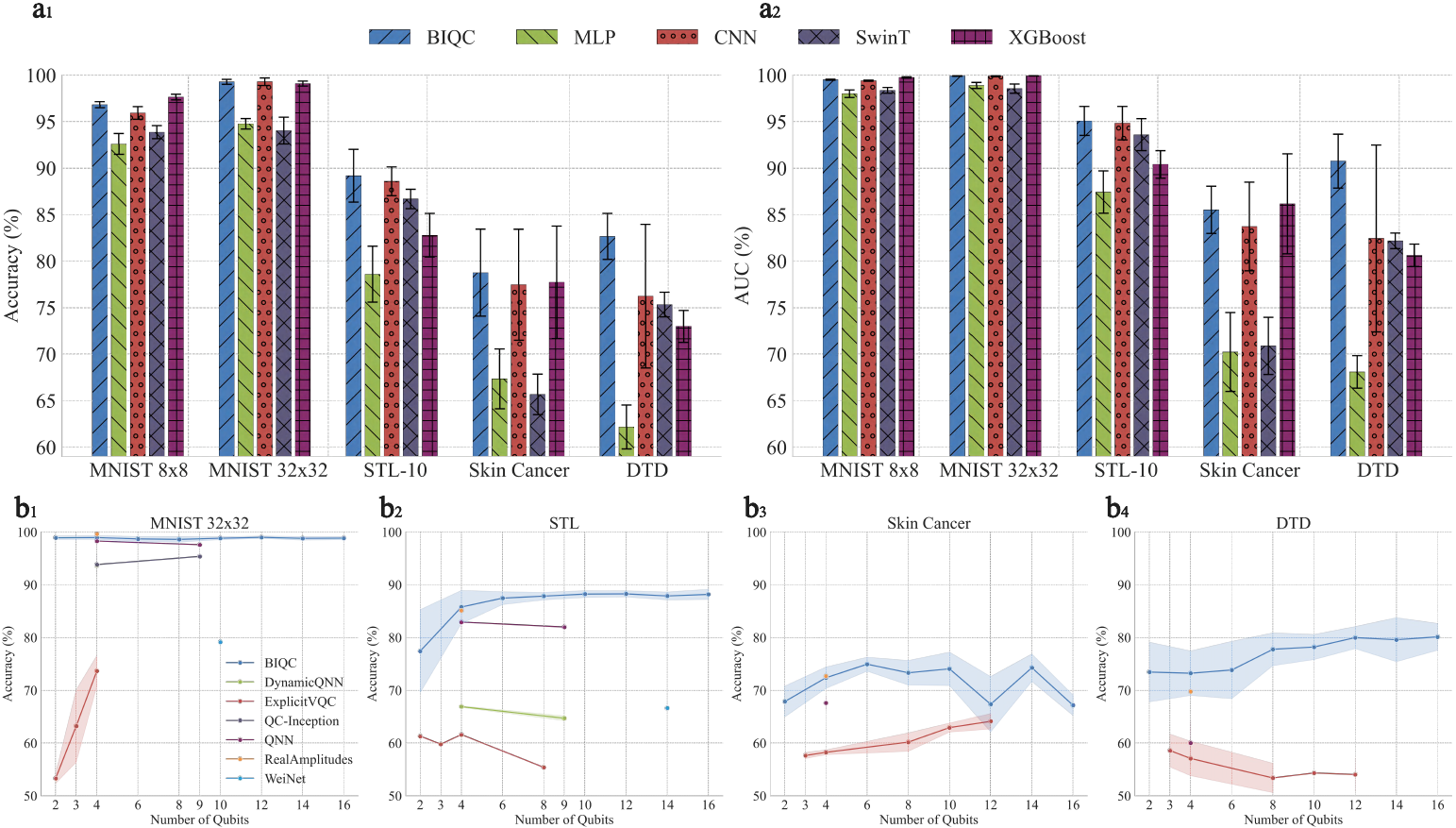}
    \caption{\textbf{Performance comparison of BIQC against classical and quantum baselines.}
    \textbf{a}, Comparison of Accuracy and AUC for BIQC against classical baselines (MLP~\cite{hinton1990connectionist}, CNN~\cite{lecun1995comparison}, SWinT~\cite{liu2021swin}, XGBoost~\cite{chen2016xgboost}) on various datasets. BIQC demonstrates consistently superior and stable performance compared to these classical methods.
    \textbf{b}, Accuracy comparison of BIQC with pure quantum network (WeiNet~\cite{ref16}), parallel hybrid quantum-classical algorithm (QC-Inception~\cite{ref31}), and serial hybrid quantum-classical methods (QNN~\cite{henderson2020quanvolutional}, RealAmplitudes~\cite{sebastianelli2021circuit}, ExplicitVQC~\cite{ref20}, DynamicQNN~\cite{ref30}) across datasets and varying numbers of qubits. Notably, BIQC's performance typically improves with increasing qubits, unlike several quantum counterparts, suggesting the Guiding Paradigm more effectively harnesses quantum resources.}
    \label{fig:results}
\end{figure*}

\section{Results and discussion}

The BIQC is evaluated on four diverse image datasets, encompassing varying resolutions and complexities.
For assessing performance across different input dimensionalities, a subset of the MNIST~\cite{lecun1998gradient} dataset consisting of digits "3" and "5" is used for binary classification, with experiments conducted at both $8 \times 8$ and $32 \times 32$ resolutions.
To evaluate performance on more complex, real-world image content, we employ the STL-10~\cite{coates2011analysis} dataset, utilizing the "airplane" and "bird" categories at its native $96 \times 96$ resolution.
The Skin Cancer~\cite{claudio2019skin} dataset, a balanced collection of $224 \times 224$ resolution dermatological images for binary classification of malignant and benign skin lesions, serves to test the model's ability to discern subtle morphological and textural details in high-resolution medical images.
Finally, the DTD~\cite{cimpoi2014describing} is used to evaluate texture classification performance, with images exhibiting "square" and "point" texture patterns selected and used at $224 \times 224$ resolution.

All datasets follow a uniform preprocessing pipeline: they are first converted to grayscale to emphasize structural and frequency features, then normalized using Min-Max scaling to the range $[-\frac{\pi}{2}, +\frac{\pi}{2}]$ to enhance training stability.

Model training and simulations use Python 3.10, PyTorch 2.5, and PennyLane 0.38 on Ubuntu 20.04 LTS.
Model parameters are optimized with the AdamW optimizer.
Training is performed with a learning rate of $1 \times 10^{-3}$ and a weight decay of $5 \times 10^{-4}$ across all datasets.
An early stopping strategy is employed, terminating training when the loss function plateaus, or after a maximum of 10,000 epochs.

Binary cross-entropy is used as the loss function for all experiments, and performance is evaluated using Accuracy and AUC metrics.
The HSF patch size is set by default to $4 \times 4$ pixels for input images with a resolution of $32 \times 32$ pixels or lower, and $32 \times 32$ pixels for higher-resolution inputs.

To assess the robustness of BIQC to noise, we introduce noise into the quantum simulator by applying $R_x(\theta)$ to each qubit, where $\theta \sim U(0, \text{noise})$.

\subsection{Benchmark performance evaluation}

Four datasets collectively challenge the scalability, resolution, and complexity dimensions, thereby providing a robust testbed for benchmarking hybrid quantum-classical methods (Fig.~\ref{fig:results}).

BIQC demonstrates consistently superior performance across all datasets compared to both classical networks and established QML algorithms.
Notably, BIQC achieves substantial gains on the challenging, high-resolution DTD dataset, outperforming the top baseline (CNN) by a significant margin of $8.42\%$ in accuracy and $10.08\%$ in AUC.
This highlights BIQC's enhanced ability to discern subtle textural variations critical for complex texture recognition tasks.
Furthermore, on STL-10, BIQC attains $89.20\%$ accuracy and $95.08\%$ AUC, surpassing competing approaches by leveraging its dual-pathway design for detailed high-frequency feature extraction.

As shown in Fig.~\ref{fig:results}(b), existing quantum methods like WeiNet, QC-Inception, and QNN face fundamental scalability challenges with high-dimensional inputs.
For instance, classifying a $224 \times 224$ image quantumly requires WeiNet a minimum of $\lceil \log_2 (224^2) \rceil = 16$ qubits via amplitude encoding, whereas BIQC achieved strong performance using only 6 qubits.
Similarly, quantum convolutional methods like DynamicQNN and QNN, while using small kernels ($4$ or $9$ qubits), must operate across the $224 \times 224$ full image.
This could demand $(224/32)^2 = 49$ times the quantum operational load compared to BIQC processing only a $32 \times 32$ patch.
This substantial resource efficiency stems directly from BIQC's Guiding Paradigm, which addresses the scalability limitations imposed by increasing input dimensions, enabling effective processing of both low- and high-resolution images.
Furthermore, BIQC exhibits a more stable and generally improving performance trend with increasing qubits compared to quantum baselines like ExplicitVQC.
This stable and scalable performance profile indicates that the Guiding Paradigm represents a more robust and efficient way to leverage quantum resources.

Furthermore, BIQC demonstrates superior image recognition accuracy, particularly when processing high-resolution and detail-sensitive datasets.
Its quantum channel adeptly extracts crucial HSF features, such as edges, textures, and fine morphological variations, essential for tasks like skin lesion recognition or texture differentiation.
This capability arises from the exponential scaling in feature representation possible with data re-uploading quantum circuits, unlike classical methods where HSF capacity scales linearly~\cite{ref21}.
This targeted quantum processing enables high performance, exemplified by $78.76\%$ accuracy and $85.53\%$ AUC on the Skin Cancer dataset for subtle pattern recognition.
Crucially, by confining quantum computation to localized HSF regions, BIQC improves scalability for large images and reduces demands on quantum hardware, offering a effective route towards leveraging quantum resources effectively.

The pronounced gains achieved by BIQC across resolution scales underscore the efficacy of the Guiding Paradigm for hybrid quantum-classical networks.
By using classical pathways for global features and targeting quantum circuits at complex local regions, BIQC overcomes scalability bottlenecks found in previous QML frameworks.
This combination of low- and high-frequency processing enhances classification accuracy while keeping quantum resource use manageable, even in high-resolution tasks.
Collectively, these results demonstrate the potential of brain-inspired, dual-channel designs to leverage quantum advantages in computer vision challenges.

\begin{figure}
   \includegraphics[width=0.95\linewidth]{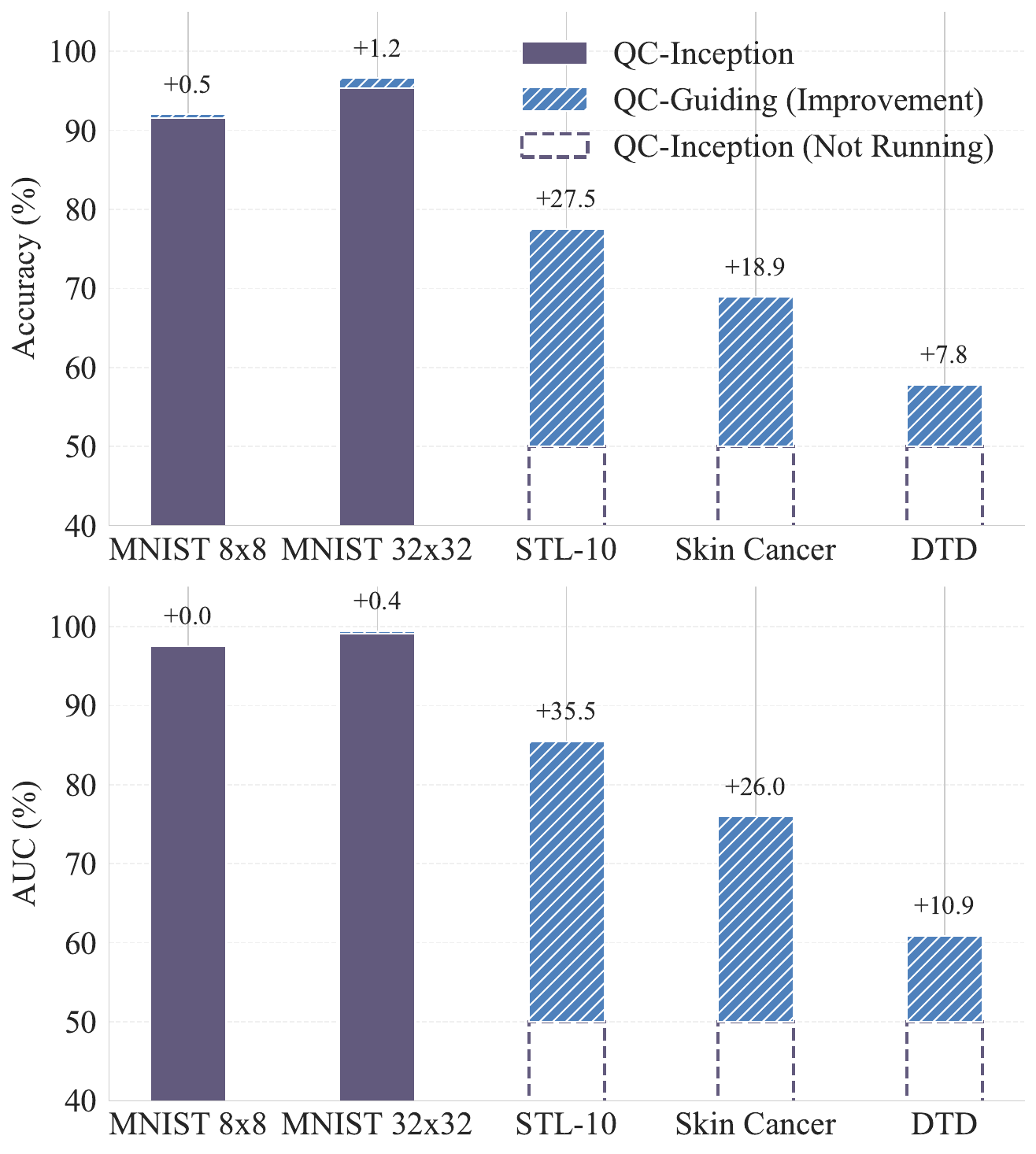}
   \caption{\textbf{Universality of the Guiding Paradigm.} Accuracy and AUC comparison for QC-Inception and its guided version, QC-Guiding.
Original QC-Inception, which requires quantum convolution operations across the entire input image, faces scalability challenges with higher-dimension datasets like STL ($96 \times 96$), Skin Cancer ($224 \times 224$), and DTD ($224 \times 224$).
By leveraging the Guiding Paradigm to concentrate quantum convolutions on localized patches of the original input, QC-Guiding effectively overcomes this input dimension barrier, enabling the processing of high-resolution datasets.}
   \label{fig:guiding}
\end{figure}

\subsection{Universality of the guiding paradigm}

The Guiding Paradigm greatly enhances the capabilities of existing QML methods.
Original QC-Inception faces scalability challenges with higher-dimension datasets because it requires quantum convolution operations across the entire input image.
As shown in Fig.~\ref{fig:main}(b), QC-Inception is limited to processing only low-resolution images such as MNIST $8 \times 8$ due to the parallel paradigm constraints.
Integrating the Guiding Paradigm's 2D-DFT-based HSF region identification enables its guiding version, i.e., ``QC-Guiding'' to overcome input dimension limitations.
QC-Guiding effectively processes high-resolution inputs up to $224 \times 224$, demonstrating the paradigm's ability to target essential local patches in large-scale inputs (Fig.~\ref{fig:guiding}).
This universality indicate the Guiding Paradigm's potential to usher in a new era of hybrid QML, characterized by broadly scalable and efficient methods capable of addressing complex, high-dimensional data challenges.

\begin{figure}
   \includegraphics[width=0.95\linewidth]{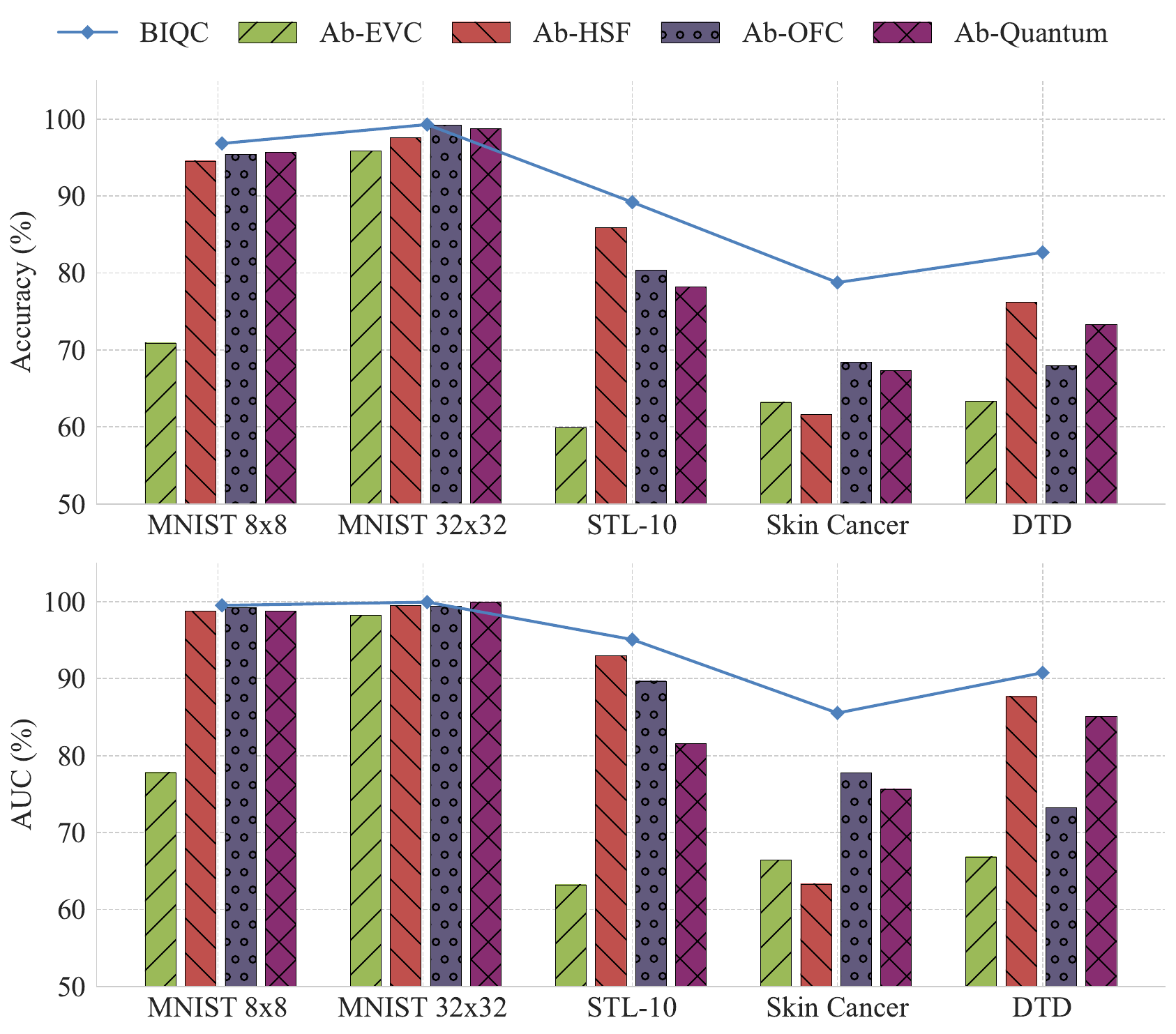}
   \caption{\textbf{Ablation study of BIQC components.} Performance impact of ablating key modules of the BIQC.
   Ablations include: (Ab-EVC) remove the EVC module, (Ab-OFC) disable OFC attention-based HSF localization, (Ab-HSF) remove the HSF pathway entirely, and (Ab-Quantum) replace the quantum circuit in the Fusiform with a classical MLP.}
   \label{fig:ablation}
\end{figure}

\subsection{Analysis of component contributions}

To investigate the contributions of each modules in BIQC and validate their correspondence to the high- and low-frequency mechanisms, we conducted ablation experiments.
Fig.~\ref{fig:ablation} summarizes the results, quantifying the performance impact of removing key BIQC components across diverse datasets.
The results demonstrate that the EVC guarantees fundamental accuracy, the quantum component excels with high-resolution details, while the HSF pathway and OFC localization are vital for complex medical and texture images.

Removing the EVC module, which eliminates low-frequency feature extraction, significantly reduces accuracy on MNIST $8 \times 8$.
This underscores the foundational role of global shape and contour recognition, similar to the EVC's role in quick visual processing.
Eliminating the HSF pathway, leaving only the LSF pathway, also lowers accuracy, confirming the benefit of integrating both coarse and fine-grained features, analogous to the complementary LSF-HSF channels in human vision.
Disabling OFC attention-based HSF localization reduces performance on high-resolution datasets, emphasizing the need to accurately target HSF regions to utilize quantum benefits.
Finally, replacing the data re-uploading quantum circuit in the Fusiform module with a classical MLP results in a performance decline on DTD, indicating that quantum processing is crucial for capturing intricate high-frequency texture features, emulating the Fusiform's detailed HSF analysis.

In summary, these ablation studies validate the functional necessity of each component.
Their synergistic integration is crucial for effective image recognition across diverse and complex datasets, which mimicked the human visual system's architecture.
LSF processing through the EVC ensures control over the global image structure, while targeted identification of HSF regions and leveraging data re-uploading enable BIQC to analyze high-frequency local features with greater precision.

\begin{figure*}[!t]
    \centering
    \includegraphics[width=1.0\linewidth]{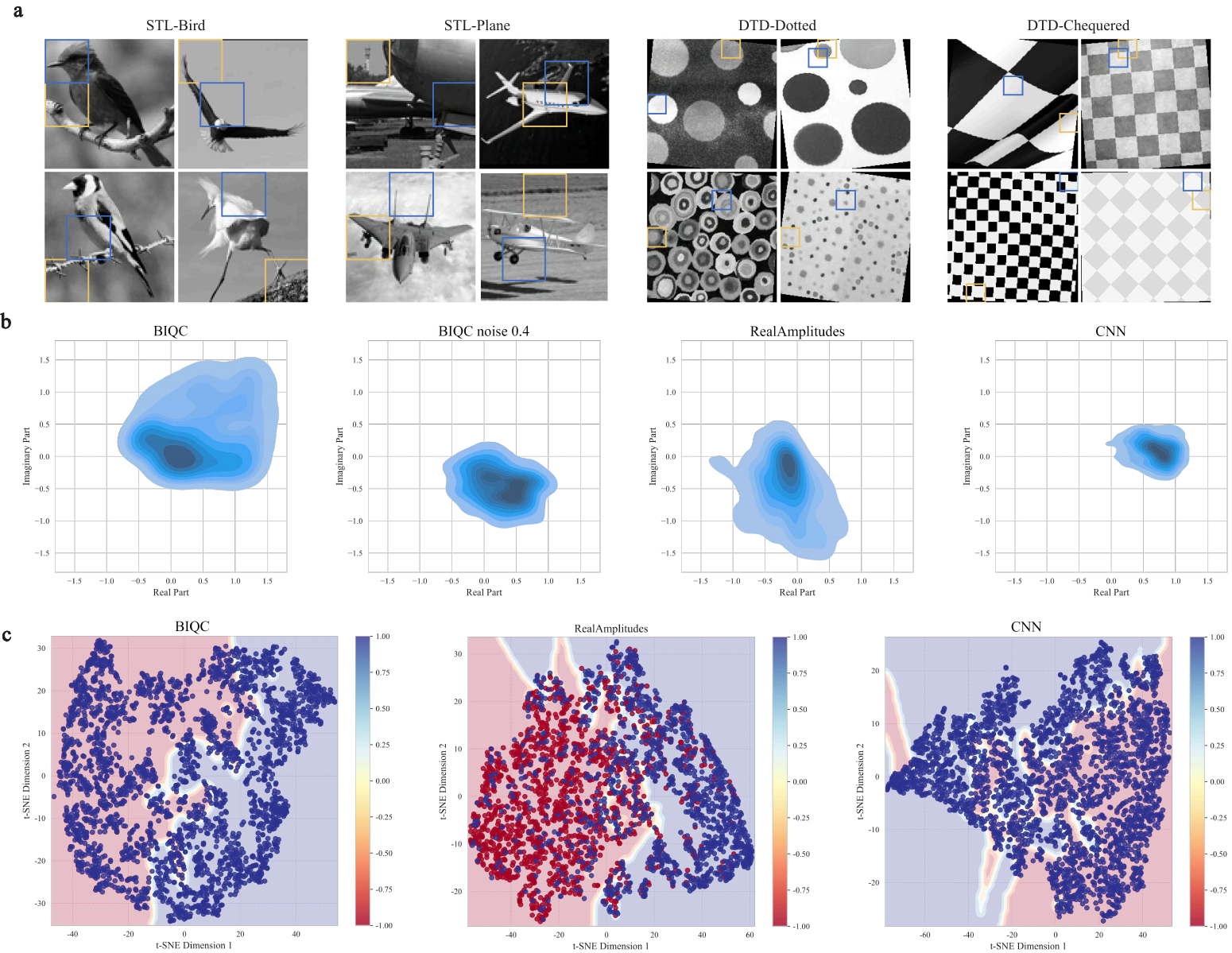}
    \caption{
    \textbf{Visualizing the Guiding Paradigm and Feature Representations of BIQC.}
    \textbf{a}, Illustration of the Guiding Paradigm's region selection.
    On the STL dataset, the blue box (simulating human attention via OFC) tends to focus on regions discriminative between classes, such as a bird's head or an airplane's wing or landing gear.
    The yellow box (driven by the 2D-DFT complexity metric via VVS) consistently targets areas with the most intricate local details; while often crucial for classification, it can occasionally focus on irrelevant complex background details (e.g., the tower in the upper left subplot of STL-Plane).
    On the DTD dataset, both strategies typically localize edges of dots or squares beneficial for distinguishing textures.
    These two strategies complement each other, guiding the quantum ansatz towards the most critical image regions.
    \textbf{b}, Fourier spectrum visualization on the Skin Cancer dataset.
    Quantum methods like BIQC and RealAmplitudes exhibit broader Fourier spectra compared to classical baselines.
    The introduction of noise influences BIQC's spectrum distribution.
    \textbf{c}, t-SNE~\cite{van2008visualizing} visualization of feature space on the Skin Cancer dataset.
    BIQC demonstrates better class separability and smoother boundaries between clusters compared to baselines.
    }
    \label{fig:visualization}
\end{figure*}

\subsection{Visualization of operational mechanisms}

To gain insight into BIQC's operational mechanisms and feature learning, we employ a multi-faceted visualization approach.
This includes visualizing the Guiding Paradigm's region selection process, analyzing high-frequency feature expressiveness through Fourier spectra, and assessing the learned feature representations using t-SNE embeddings.

Fig.~\ref{fig:visualization}(a) visualizes the complex regions identified by the Guiding Paradigm from raw input images.
As observed, the blue, human-like attention regions tend to focus on critical patches within the STL dataset that are discriminative between bird and plane.
In contrast, the yellow regions, derived from the complexity metric, are more inclined to pinpoint the most locally complex areas within the image.
For instance, in the last STL-Bird subplot, the VVS identifies a region encompassing avian digits and arboreal bark.  
However, relying solely on complexity-based region localization is insufficient, as exemplified in the first STL-Plane subplot, where the metric erroneously targets distant buildings in the image background, which are irrelevant to the classification task.
These two approaches for local region localization, while significantly reducing quantum computing resource requirements, complementarily guide the quantum circuit towards the most classification relevant regions of the image, which is at the heart of BIQC's efficiency and effectiveness.

To examine high-frequency expressiveness, we perform Fourier analysis on hidden layer activations, computing Fourier coefficients at a normalized frequency $\omega = 1.0$, corresponding to the Nyquist frequency.
Fig.~\ref{fig:visualization}(b) visualizes the kernel density estimation of these coefficients.
Observations reveal that hybrid methods exhibit moderately dispersed Fourier spectra, suggesting a capability in representing a range of HSF.
Notably, BIQC, especially with noise, shows a spectrum distribution that is dispersed yet more concentrated than RealAmplitudes, potentially indicating a balanced representation of relevant high-frequency information.
This aligns with the Guiding Paradigm's emphasis on effective HSF capture.

We employ t-SNE to visualize the learned feature spaces of BIQC, RealAmplitudes, and CNN on the Skin Cancer dataset (Fig.~\ref{fig:visualization}(c)).
t-SNE projections show that while CNN displays class intermixing, both hybrid quantum-classical models (BIQC, RealAmplitudes) exhibit better-defined clusters.
Crucially, BIQC demonstrates more refined class separation and less cluster intermingling compared to RealAmplitudes.
This improved feature space organization supports the effectiveness of the Guiding Paradigm in enabling BIQC to capture subtle, diagnostically relevant features in complex medical images.

\section{Summary and outlook}

This study demonstrates the efficacy of a novel brain-inspired approach for hybrid QML in high-dimensional image classification.
Our proposed algorithm, BIQC, consistently outperforms both classical baselines and existing QML methods across a diverse range of datasets, including challenging high-resolution and texture-rich examples like Skin Cancer and DTD.
As evidenced by quantitative results and visualizations (Fig.~\ref{fig:results}, Fig.~\ref{fig:visualization}), BIQC effectively overcomes the scalability limitations inherent in prior parallel or serial quantum-classical frameworks~\cite{ref31, henderson2020quanvolutional, sebastianelli2021circuit, ref35, ref36, ref22, ref38} by strategically allocating computational resources.
This enhanced performance stems from two core innovations.

(i) The Guiding Paradigm represents a conceptual shift, moving away from processing entire inputs with quantum circuits towards a targeted application of quantum computation.
By first employing classical methods to analyze global low-frequency information and subsequently identify complex, potentially discriminative high-frequency regions, the paradigm directs quantum resources only where they are most needed.
This localized quantum processing drastically reduces the input dimensionality handled by the quantum circuit, thereby mitigating qubit and quantum operation requirements, crucial for near-term hardware~\cite{mcclean2018barren, cerezo2022challenges}.
The visualization results (Fig.~\ref{fig:visualization}(a)) confirm that the paradigm effectively identifies relevant image patches, using complementary attention and complexity metrics.
Furthermore, the paradigm's universality is highlighted by its ability to enhance existing QML methods like QC-Inception, enabling them to process high-resolution inputs previously intractable (Fig.~\ref{fig:guiding}).

(ii) Central to BIQC's success is the Complementarity Architecture, drawing inspiration from the human visual system's dual HSF and LSF processing streams~\cite{ref23, ref24, ref25, ref26}.
This architecture assigns distinct roles: a classical pathway efficiently processes global LSF information (contours, shapes), while a quantum pathway focuses on analyzing localized HSF patches (textures, edges, fine details) using data re-uploading circuits.
This division harnesses the strengths of both modalities, combining classical efficiency for large-scale structure and the potential representational power of quantum circuits for complex, high-frequency patterns, potentially offering scaling advantages in feature representation relative to circuit size~\cite{ref21}.
Ablation studies (Fig.~\ref{fig:ablation}) rigorously validate this synergy, demonstrating that removing either the LSF or HSF path significantly degrades performance, confirming the necessity of integrating both coarse and fine-grained feature analysis.

Future work includes validating BIQC on NISQ hardware while refining its components, with a focus on optimizing the identification of critical complex regions.
Furthermore, the paradigm's demonstrated universality points towards broader applications in scalable hybrid models for complex, high-dimensional data across various scientific and industrial domains.

\section{Acknowledgement}
We thank Peixin Shen for his valuable discussions and suggestions for modification to this manuscript.
This work was supported by the National Science Fund for Distinguished Young Scholars of China (Grant No. 62325601).

\bibliographystyle{quantum}
\bibliography{bib}

\end{document}